# *Detection of Diabetic Anomalies in Retinal Images using Morphological Cascading Decision Tree.*

Faisal Ghaffar[1], Sarwar Khan[2], Bunyarit Uyyanonvara[1], Chanjira Sinthanayothin[3] and Hirohiko Kaneko[4]

[1]School of ICT, Sirindhorn International Institute of Technology [SIIT], Thammasat University, Thailand.
[2]Faculty of Engineering, kasetsart University, Thailand.
[3]National Electronics and Computer Technology Centre [NECTEC], Thailand.
[4]Tokyo Institute of Technology, Ookayama Campus, Meguro-ku, Tokyo, Japan.

Email: engr.faisal90@gmail.com, say2sarwar@gmail.com, bunyarit@siit.tu.ac.th, chanjira.sinthanayothin@nstda.or.th and kaneko@ip.titech.ac.jp, respectively

## Abstract

**This research aims to develop an efficient system for screening of diabetic retinopathy. Diabetic retinopathy is the major cause of blindness. Severity of diabetic retinopathy is recognized by some features, such as blood vessel area, exudates, haemorrhages and micro aneurysms. To grade the disease the screening system must efficiently detect these features. In this paper we are proposing a simple and fast method for detection of diabetic retinopathy. We do pre-processing of grey-scale image and find all labelled connected components (blobs) in an image regardless of whether it is haemorrhages, exudates, vessels, optic disc or anything else. Then we apply some constraints such as compactness, area of blob, intensity and contrast for screening of candidate connected component responsible for diabetic retinopathy. We obtain our final results by doing some post processing. The results are compared with ground truths. Performance is measured by finding the recall (sensitivity).
We took 10 images of dimension 500 * 752. The mean recall is 90.03%.**

## A. Introduction

Diabetes is caused when pancreas does not produce insulin properly or response of body cells is not adequate to insulin. This directly affects blood vessels in body. In retina it results in exudates, haemorrhages and micro aneurysms. Leakage of fats or lipids from abnormal blood vessels forms exudates. Blood leakage from vessels causes Haemorrhages. Micro aneurysms appear as small round dark red dots. These abnormalities in eye are termed as diabetic retinopathy. It can lead to blindness if not diagnosed properly and timely. The potential risk of blindness is decreased by 50% if patients are screened for development of diabetic retinopathy [1-3].

Many techniques are used for the detection of these abnormalities. Thomas Walter et al. first found candidate regions; these are regions that possibly contain exudates. Then applied morphological techniques to find the exact contours[4]. D. Kavitha. S. Shenbaga Devi segmented blood vessels using median filtering and found the convergent point (CP) of blood vessels using the least square fitting then extracted brighter regions using thresholding then combined both, determined the location of optic disc and classified the exudates[5]. Akara Sopharak et.al used fuzzy c-mean clustering technique to obtain clusters from pre-processed retinal image and then applied four morphological features intensity, standard deviation, hue and number of edge pixels to classify exudate clusters from non-exudates [6]. Chanjira sinthanayothin et.al used a combinational approach by first obtaining pre-processed image and defined two main features optic disc and vessels. Exudates are obtained by recursive region growing and haemorrhages and micro aneurysms are obtained with same features as that of vessels [7]. Saiprasad Ravi Shankar et.al detects proposes a new constraint for optic disk detection. They first detected the major blood vessels and then use the intersection of these to find the approximate location of the optic disk. Exudates are extracted using open and close operations using filters of different sizes While micro aneurysms and haemorrhages are segmented using morphological filters that exploit their local 'dark patch' property[8]. Akara Sopharak et.al detected exudates by first eliminating optic disc and then detected exudates through morphological filters [9]. Akara Sopharak et.al used naive Bayes classifier for detection of exudates [10].

In this paper we are proposing a method to develop a technique which will detect exudates, haemorrhages and micro aneurysms in retina. We first do pre-processing of image to remove noise and enhance some features of image then find all the connected components in the image. We try to obtain as many blobs as we can an image. Then we apply morphological filters on some features such as area, compactness, contrast intensity and hue to detect the candidate regions. This method is quite simple because we only apply few numbers of morphological filters to pre-processed blobs of an image and detect the suspected anomalies.

This paper is further classified as section B explains the whole methodology. The sub section 1 of methodology explains pre-processing of the original image; sub section 2 gives details about the blobbing of image; sub section 3 explains the morphological threshold applied to obtain the candidates; sub section 4 explains about post-processing. Section C gives detail about the result of tested dataset and finally the section D concludes the whole technique.

## B. Methodology

**1. Pre-processing:** The purpose of the pre-processing is to remove noise and obtain maximum blobs from the image regardless of whether its candidate area or not. We take colour image resize it to standard size and convert it to grey scale image shown in fig 1(a) Then we applied adaptive histogram equalization for enhancement of contrast (CLAHE) shown in fig 1(b). As we have to obtain both bright and dark regions from the image so we use different approaches to obtain all. To obtain bright regions from image we apply morphological top hat filtering and bottom hat filtering with disk shape of radius 12. Subtraction of bottom hat filtering from top hat filtering gives us all the bright spots in the image shown in fig 1(c). Adjusting the intensity and then applying morphological closing and opening gives us refine result of bright regions shown in fig 1(d). We call this image suspected exudates image (SEI). To obtain dark spots in image we again apply top hat and bottom hat filtering with disk shape of radius of 12. Subtraction of top hat from bottom hat gives us all the dark regions in image shown in fig 1(e). The contrast is further enhanced and then morphological closing and opening is applied to the image to get a refined result shown in fig 1(f). We call this image as suspected haemorrhages image (SHI). Further processing is separately on these two SEI and SHI.

**2. Blobbing of image:** The pre-processing gives us two binary images. Region based segmentation is applied to both SEI and SHI images separately and all the connected components each image are labelled as single blob.

**3. Morphological filters:** After labelling all the components in the image we filtered blobs on basis of Area, compactness, intensity and hue. A cascading decision tree shown in fig 2 is used for filtering out most of the non- candidate blobs.

Area of blob is calculated by calculating actual number of pixels in that area. The area filter removes the blobs having very large area or very small. Blobs having pixels less than 5 and greater than 5000 were marked as non-candidate blobs in area filter. Remaining blobs after applying area filter are shown in fig 3(a). Most of the vessels are removed by applying compactness filter. The blobs having vary high compactness is marked as non-candidates. The minimum and maximum threshold values for SEI image are 0.55 and 9 respectively. While for SHI image is 0.7 and 4 respectively. The blobs which fall outside these thresholds were marked as non-candidate blobs. The candidate blobs after applying compactness filter is shown in fig 3(b). Haemorrhages have low intensity value and exudates have high intensity value [11]. We remove blobs having low mean intensity value from SEI and blobs having high mean intensity value from SHI. For each blob average of the maximum and minimum intensities were found. For SEI image all the blobs having average value less than 90 were marked as non-candidate and for SHI image all the blobs having average value greater than 200 were marked as non-candidate. Resultant blobs after applying intensity filter is shown in fig 3(c). Lastly we apply threshold value to remove blobs having high hue or low hue. Mean hue value of each blob was found. The blobs having mean hue value less than 0.125 and mean value higher than 0.165 in SEI image and mean value less than 0.06 and higher than 0.125 were marked as non-candidate blobs. Resultant blobs after applying hue filter is shown in fig 3(d).

All these thresholds were set so that we should minimize the number of blobs but it should not remove the candidate blobs. These values were after quite lot of experiments on different thresholds.

**4. Post-Processing:** Results from SEI and SHI images are combined. Centre and orientation of each blob is found. Finally we draw an ellipse according to the orientation around the blob. The ellipse is beneficial than circle because if the blob has stretched shape, the circle misses some parts of the blob while the ellipse surround the whole blob. The resultant image is shown in fig 4.

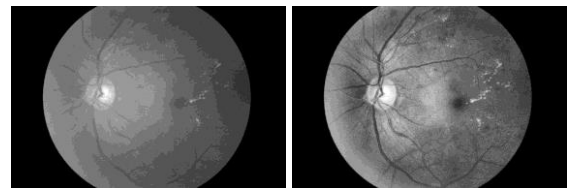

**1( a ). Grey-scale image. 1( b ). Enhanced image**

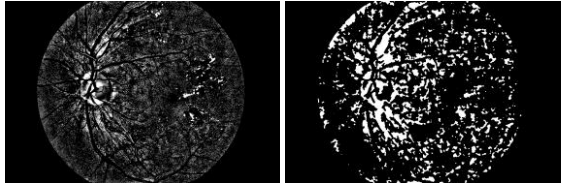

**1( c ).** Top hat filtering of enhanced image.  **1( d ).** Morphological opening and closing.

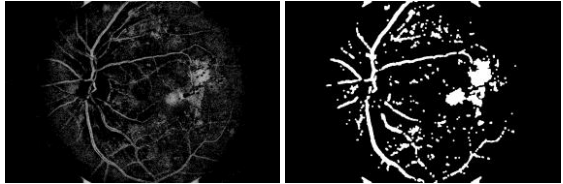

**1( e ).** Bottom hat filtering of enhanced image.  **1( f ).** Morphological opening and closing.

**Figure 1: Pre-processing**

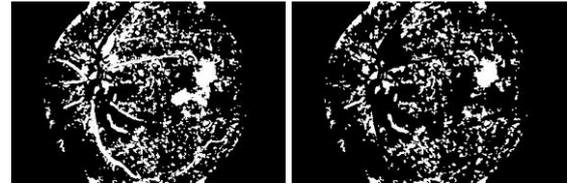

**3( a ).** Area filter  **3( b ).** Compactness filter

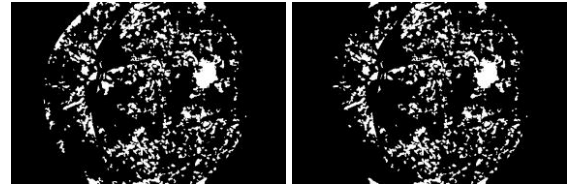

**3( c ).** Intensity filter  **3( d ).** Hue filter

**Figure 3: Combine Result of SEI and SHI after applying each case of Decision Tree**

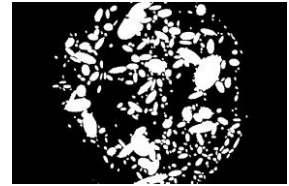

**4. Post processing: Ellipse drawn on each blob.**

### C. Result Evaluation

The final results were checked against ground truths. The results are pixels to pixels comparison of ground truths and our result. Also our focus is on blobs and we extract a large number of blobs from our image so we used a limited dataset of 10 images in result evaluation. We calculated true positive (TP) and true negative (TN) values. The evaluation technique is pixel to pixel comparison because we are focusing only on the sensitivity of this method. To check the performance we calculated the Recall (Sensitivity) of all images.

$$\text{Sensitivity} = \frac{\text{TP}}{\text{TP} + \text{FN}}$$

We finally calculated the mean recall of all the 10 images. The final mean recall is 94.5%. Table 1 shows the number of blobs and recall after Pre-processing, each step of decision tree and post-processing.

### D. Conclusion and Future work

This paper concludes that using morphological cascading tree for filtering candidate blobs from pool of blobs has positive impact on overall process

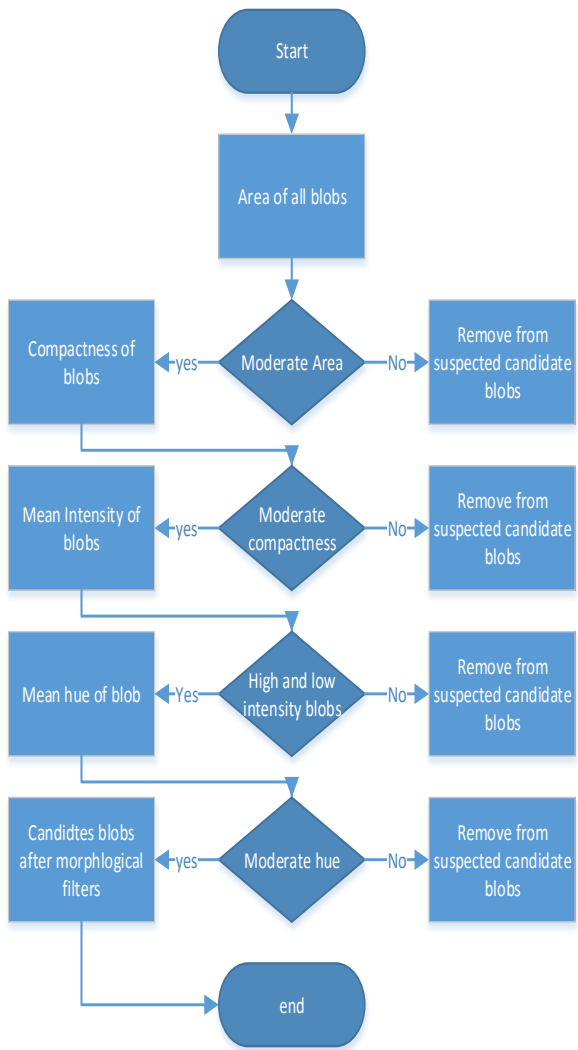

**Figure 2. Cascading Decision Tree**

of detection of anomalies in retinal images. The final results have high recall values.

The future work includes improvement of morphological cascading tree, its comparison with other morphological techniques on the basis of its accuracy and correctness and also using machine learning approaches to increase the recall as well sensitivity of the whole technique.

**Acknowledgement**

This research is financially supported by Thailand Advanced Institute of Science and Technology (TAIST), National Science and Technology Development Agency (NSTDA), Tokyo Institute of Technology, Sirindhorn International Institute of Technology (SIIT), Thammasat University (TU) under the TAIST Tokyo Tech Program.

| Image no. | After Pre-processing | | After Area Filter | | After compactness filter | | After intensity filter | | After hue filter | | After post-processing | |
|---|---|---|---|---|---|---|---|---|---|---|---|---|
| | Total blobs | Recall % | Total blobs | Recall % | Total blobs | Recall % | Total blobs | Recall % | Total blobs | Recall % | Total blobs | Recall % |
| 1. | 830 | 100 | 684 | 100 | 495 | 72.5 | 442 | 72.5 | 433 | 72.5 | 253 | 93.46 |
| 2. | 654 | 90.8 | 518 | 90.4 | 334 | 85.0 | 303 | 84.8 | 280 | 84.8 | 190 | 87.11 |
| 3. | 827 | 100 | 693 | 100 | 504 | 97.2 | 411 | 96.0 | 408 | 96.0 | 202 | 91.56 |
| 4. | 772 | 100 | 670 | 99.7 | 507 | 88.6 | 436 | 88.3 | 414 | 87.9 | 132 | 81.85 |
| 5. | 841 | 100 | 687 | 100 | 486 | 95.5 | 473 | 95.5 | 471 | 95.5 | 261 | 94.54 |
| 6. | 804 | 99.2 | 634 | 96.7 | 432 | 9.62 | 424 | 96.2 | 424 | 96.2 | 225 | 91.40 |
| 7. | 838 | 100 | 690 | 100 | 501 | 100 | 205 | 86.5 | 197 | 86.5 | 97 | 85.25 |
| 8. | 851 | 99.6 | 691 | 99.6 | 485 | 97.0 | 470 | 97.0 | 442 | 97.0 | 234 | 93.34 |
| 9. | 721 | 100 | 586 | 96.9 | 413 | 96.9 | 409 | 96.9 | 409 | 96.9 | 240 | 93.13 |
| 10. | 675 | 99.6 | 556 | 79.5 | 359 | 78.1 | 357 | 78.1 | 355 | 78.1 | 227 | 88.74 |
| Mean | 781 | 98.9 | 641 | 96.3 | 452 | 90.7 | 346 | 89.2 | 383 | 89.1 | 206 | 90.03 |

**Table 1. Comparison of number of blobs and recall after each case of cascading decision tree.**